\documentclass[usenatbib]{mn2e}
\input{epsf}
\usepackage{times}
\usepackage{epsfig}
\usepackage{graphicx}  
\usepackage{dcolumn}   
\usepackage{bm}        
\usepackage{subfigure}
\usepackage{amsmath, amssymb}

\def\etal{\rm et al~}

\begin{document}

\title[]{Generalizing thawing dark energy models: the standard vis-\`{a}-vis model independent diagnostics}

 
\author[]{Debabrata Adak$^1$
, Debasish Majumdar$^1$
and Supratik Pal$^2$
\\$^1$Astroparticle Physics and Cosmology Division, Saha Institute of Nuclear Physics, 1/AF Bidhannagar, Kolkata 700064, India\\
$^2$Physics and Applied Mathematics Unit, Indian Statistical Institute, 
203 B.T.Road, Kolkata 700108, India}


\maketitle

\begin{abstract}
We propose a two parameter generalization for the dark energy
equation of state (EOS) $w_X$ for thawing dark energy models which
includes PNGB, CPL and Algebraic thawing models as limiting
cases and confront our model with the latest observational data namely  
SNe Ia, OHD, CMB, BOSS data.  
Our analysis reveals that the phantom type
of thawing dark energy is favoured 
upto $2 \sigma$ confidence level. 
These results also show that 
thawing dark energy EOS is not unique
from observational point of view. Though different 
thawing dark energy models are not distinguishable 
from each other from best-fit values 
(upto $2\sigma$ C.L.s) of matter 
density parameter ($\Omega_m^0$) and hubble parameter ($H_0$) at present epoch, 
best-fit plots of linear growth of matter 
perturbation ($f$) and average deceleration parameter ($q_{\rm av}$);
the difference indeed reflects in best-fit variations  
of thawing dark energy EOS, model-independent geometrical diagnostics like
the statefinder pair $\{r,~s\}$ and $Om3$ parameter. We are thus led to 
the conclusion that unlike the standard observables ($\Omega_m^0$,
$H_0$, $f$, $q_{\rm av}$), the model-independent parameters 
($r,~s,~Om3$) and the variations of EOS (in terms of $w_X-w_X'$ plots) serve as
model discriminators for 
thawing dark energy models.

\end{abstract}

\section{Introduction}
\label{intro}

Late time cosmic acceleration at the present epoch has almost
been 
a de facto phenomenon since the late nineteens. 
Advances in cosmological observations during the past two decades 
reveal strong evidences in favour of this accelerated expansion of the
universe.
These evidences have been
brought forth 
{\em \`{a} la} independent astrophysical 
observations like Supernovae Type Ia (SNe Ia) luminosity 
distance modulus as a function of redshift 
\cite{sn:3f, sn:3e, sn:3b, sn:3g, sn:3j, sn:3d, sn:3c, sn:3h, sn:3a, sn:3i}, 
Observational Hubble Data (OHD) \cite{ohd:4b, Abraham:13, ohd:4c, ohd:4a, stern2010, Moresco2012, ohd:4d}
, Cosmic Microwave Background (CMB) Shift Parameter  
\cite{komatsu:5d, komatsu:5c, komatsu:5a, komatsu:5b, Bennett2012}
and Baryon Oscillation Spectroscopic Survey (BOSS) Data 
\cite{bao:6b}. 
A good deal of attempts have been taken to explain
this accelerated expansion assuming the presence of 
some exotic fluid, namely 
{\it dark energy}, in huge abundances in the universe. 
Though there exists a lot of dark energy models (see for example 
\cite{ hirano11:7b, hirano11:7f, hirano11:7l, hirano11:7e, hirano11:7i, hirano11:7a, hirano11:7h,
hirano11:7j, hirano11:7m, hirano11:7c, hirano11:7d, hirano11:7g, hirano11:7k} and 
references therein) 
with standard as well as exotic ideas;
the canonical and non-canonical scalar fields are the most 
promising candidates till date. 
Of late, Robert R. Caldwell and Eric V. Linder \cite{cald:8} 
categorized these scalar field models
in two broad classes namely ``freezing" and ``thawing" dark energy,
based on the asymptotic behavior of the scalar field potential. 
In thawing models, dark energy equation of state $w_X$ initially
remains at $-1$ and deviates from $-1$ near present epoch whereas
just the opposite behavior of $w_X$ is witnessed in freezing models.

Thawing models, in which we are interested in the present article,
are broadly classified into two categories: (i) quintessence (for which $w_X$ moves to $w_X^0>-1$),
and (ii) phantom (where $w_X$ is less than $-1$). A third possibility has also been explored in 
\cite{clemson2008, quin&phanThaw:9d, quin&phanThaw:9e, quin&phanThaw:9a, quin&phanThaw:9b, 
quin&phanThaw:9c} which lead
to both quintessence and phantom behavior of $w_X$.  
In these slow-rolling scalar field models with nearly flat potential, 
initially the kinetic energy of the field is much smaller than the 
potential energy. 
This is because of the initial large Hubble damping which keeps the field
nearly frozen at $w_X=-1$ at earlier era i.e., in radiation and matter
dominated eras. Due to the expansion of the universe,
energy density of the universe decreases. After the radiation and matter
dominated eras, the field energy density becomes comparable 
to the background energy density of the universe resulting in
the deviation of the field from its frozen state, thereby leading to deviation of $w_X$ from $-1$.

Slow-roll scalar field thawing models can be characterized by
different relations between $w_X$ and the scale factor $a$ of
the universe. Some typical examples of 
 CPL parametrization
(Eq. (\ref{cpl})) \cite{Chevallier:11,linder2:10}, PNGB models (Eq. (\ref{pngb})) and Algebraic thawing 
models (Eq. (\ref{algethaw})) are included in the work by E. V. Linder \cite{Linder07:1}. The corresponding equation of state
parameterizations are respectively given by,
{\small
\begin{equation}
\frac{dw_X}{d(\ln a)} = (1+w_X)
\label{cpl}
\end{equation}
}
{\small
\begin{equation}
\frac{dw_X}{d(\ln a)} = F (1+w_X)
\label{pngb}
\end{equation}
}
{\small
\begin{equation}
\frac{dw_X}{d(\ln a)} = (1+w_X)\left(3- \frac{3-p}{1+ba^{-3}}\right)\,\,,
\label{algethaw}
\end{equation}
}
where $F$ is a parameter which is inversely proportional to the symmetry 
breaking energy scale and $p$ and $b$ are two free parameters.

In the present work, we propose a two parameter generalization for this
thawing dark energy models as
{\small
\begin{equation}
\frac{dw_X}{da} = (1+w_X)f(a)
\label{ourmodel}
\end{equation}
}
where $f(a)$ is an arbitrary function of scale factor $a$. In this article, 
we have chosen $f(a)$ as $f(a)=c/a^n$, where $c$ and $n$ are two 
arbitrary parameters. In this context we would like to mention that
choice of $f(a)$ can be made otherwise and it would be interesting
to see if there exists any observational constrain on the form of $f(a)$
which is beyond the scope of this article.
With the chosen form of $f(a)=c/a^n$ for $n=1$ and $c=1$ 
our proposal exactly overlaps with CPL 
thawing dark energy model \cite{Linder07:1}.
For $n=1$ and  $1<c<3$, our proposal leads to PNGB thawing dark
energy model \cite{Linder07:1} which 
have been studied exclusively for scalar fields dark energy with PNGB potential 
\cite{pngbmodel:14b, pngbmodel:14c, pngbmodel:14a,  pngbmodel:14d}. 
For suitable choice of the parameters $n$ and $c$, our model can 
approximately reflect Algebraic thawing \cite{Linder07:1} as well.

As it turns out, all the existing (and probably, upcoming) thawing 
dark energy models fall in this broad minimal parametrization with
different values of the parameters $n$ and $c$. So, rather than 
proposing individual models, it is quite reasonable to construct a 
minimal generic form of parametrization, analyze it and search for 
possible constraints on the parameters from present-day observations.
This is the primary objective of the present article.

Along with this view, we also draw some comparisons among the results obtained for different
values of $n$ (i.e., for $n=1$, $n=1.5$ and $n=2$) with different fixed
values of $c$ and vice-versa. We further provide justification for this
proposed generalized form of thawing dark energy model against the other
existing thawing models by comparing them with ours. Moreover, we constrain our 
model by latest Supernova Type Ia Data from Union2.1 compilation
\cite{sn:3f, sn:3e, sn:3b, sn:3g, sn:3j, sn:3d, sn:3c, sn:3h, sn:3a, sn:3i},
newly released Observational Hubble parameter Data 
\cite{ohd:4b, Abraham:13, ohd:4c, ohd:4a, stern2010, Moresco2012, ohd:4d}, Cosmic Microwave Background Shift
Parameter Data from WMAP 9 year results \cite{Bennett2012}
and the latest Baryon Oscillation Spectroscopic Survey (BOSS) data from SDSS-III Data Release \cite{bao:6b}.
For such analyses we have five parameters in total namely $c$, $n$, $w_X^0$,
$\Omega_m^0$ and $H_0$ (where $\Omega_m^0$ and $H_0$ are matter density at
present epoch normalized to critical density and Hubble parameter at present
epoch respectively). Since the value of $\Omega_r^0$ (the normalized radiation
density at present epoch) is very low we do not treat it as a parameter and
consider $\Omega_r^0=5.05 \times 10^{-5}$ \cite{Beringer:pdg:2} for numericals.
Our analysis also helps in comparing the standard diagnostics with model
independent ones, and reveals the pros and cons of each one. 

The major conclusions of the paper are as follows:

\begin{itemize}

\item 
Existing thawing dark energy models \cite{Linder07:1} can be generalized
in the form of Eq. (\ref{ourmodel}) as we have presented in this article.
Our minimal generalization of thawing dark energy models 
(Eq. (\ref{ourmodel})) with two parameters $n$ and $c$ leads to different
existing thawing models namely CPL ($n=1, ~c=1$), PNGB ($n=1,~1<c<3$)
and the Algebraic thawing (suitable choices of $n,\&~c$). 

\item Results obtained for different $n$ values (with different values of $c$)
barely differ from the observational point of view. Other way around, 
we can say that values of $n$ (with different values of $c$) can 
hardly affect the best fit values as well as the $1\sigma~\&~2\sigma$
C.L.s of matter density parameter $\Omega_m^0$ and 
Hubble parameter at the present epoch $H_0$.
Also the best fit plots for redshift evolution of average deceleration
parameter $q_{\rm av}$ and the growth of matter perturbations in terms of
evolution of growth factor $f$ with redshift $z$ (best fit plots) remain
unaffected when the values of $n$ and $c$ are altered accordingly.  
Therefore it is difficult to provide a unique dark energy EOS $w_X$ for
the thawing dark energy models as different values of $n$ with different 
values of $c$ lead to the same cosmological dynamics.


\item The best fit values and the $1\sigma~\&~2\sigma$ C.L.s of EOS 
at the present epoch $w_X^0$ does leave little trace on
model discrimination for thawing dark energy. Here we discuss the fact 
that the values of $n$ and $c$ can 
be constrained by $w_X-w_X'$ ($w_X'=\frac{d w_X}{d \ln (a)}$) plots 
\cite{cald:8} for thawing dark energy models. More importantly,
best fit $w_X-w_X'$ plots can also serve as a model discriminator for the
thawing dark energy models. The non-linear $w_X-w_X'$  plots can be 
realized for values of $n$ other than $1$ with different values of $c$.
This is an important issue as PNGB and CPL parameterizations can result only
in linear $w_X-w_X'$ plots and recent works on scalar field dark energy
models point towards the non-linear $w_X-w_X'$ plots \cite{Ali:16}.

\item Most importantly, the model-independent parameters like statefinder
pair $\{r,s\}$ \cite{new1} and the so called $Om3$ \cite{new2} parameter do play 
a crucial role 
in discriminating among different dark energy models. 
Study of these parameters in the context of our generalized thawing model,
therefore, reveals the fact that unlike the standard parameters mentioned 
in 2nd major conclusion above, 
these  parameters indeed serve as model discriminators for 
different thawing dark energy models i.e., these parameters can identify
the different values of $n$ as well as $c$ in our generalized model.

The paper is organized as follows. In the next Sec. we propose the
generalization for the thawing dark energy models and mention the standard 
as well as the model independent parameters. The Sec. 3 briefly 
describes the various observational data we used. In the Sec. 4
we present the results obtained by the analyses of the various 
observational data. In the Sec. 5 we discuss our results and put
forward the conclusions of the present work. 

\end{itemize}
%
%




\section{The scheme of generalization}

\subsection{Generalized thawing dark energy EOS}


We propose a minimal two parameter generalization for thawing dark 
energy EOS $w_X$ as,
\begin{equation}
\frac{dw_X}{da} = (1+w_X)f(a)
\label{generalthaw}
\end{equation}
where $f(a)$ is an arbitrary function of scale factor $a$ of the universe.
We study the dynamical universe with radiation, matter and thawing
dark energy obeying the proposed EOS $w_X$ with $f(a)=\frac{c}{a^n}$.
The proposed choice
of $f(a)$ here, for the generalized thawing model is motivated by the following findings: 

i) for $n=1$ and $c=1$, our model is exactly same as CPL parametrization (Eq \ref{cpl}). 

ii) for $n=1$ and $c=F$ (F being the parameter described in Sec. 1),
our model is exactly same as PNGB model (Eq \ref{pngb}). 

iii) Algebraic thawing 
case  (Eq \ref{algethaw}) can also approximated for certain choices of $c$
and $n$ in terms of $b$ and $p$. 

iv) for values of $n$ other than $1$, generalized thawing dark energy
EOS takes the form

{\small
\begin{equation}
w_X(a) = -1 + (1+w_X^0)\exp\left[\frac{c}{(n-1)}(1-a^{(1-n)})\right]
\label{generalthaw2}~~~~~~~~~(n \neq 1)\,\,,
\end{equation}
}
where $w_X^0$ is the value of $w_X$ at the present epoch.
Expansion of $w_X(a)$ about $a=1$ gives,

{\small
\begin{eqnarray}
w_X(a)&=& w_X^0 - c(1+w_X^0)(1-a)
+\frac{1}{2} (1+w_X^0) (c^2-cn)(1-a)^2 \nonumber\\&&
+ \,\,{\rm higher}\,\,{\rm order}\,\,{\rm terms}\,\,.
\label{generalthaw2a}
\end{eqnarray}
}


%
\begin{figure*}
\centering
\includegraphics[height=2.5in,width=6in]{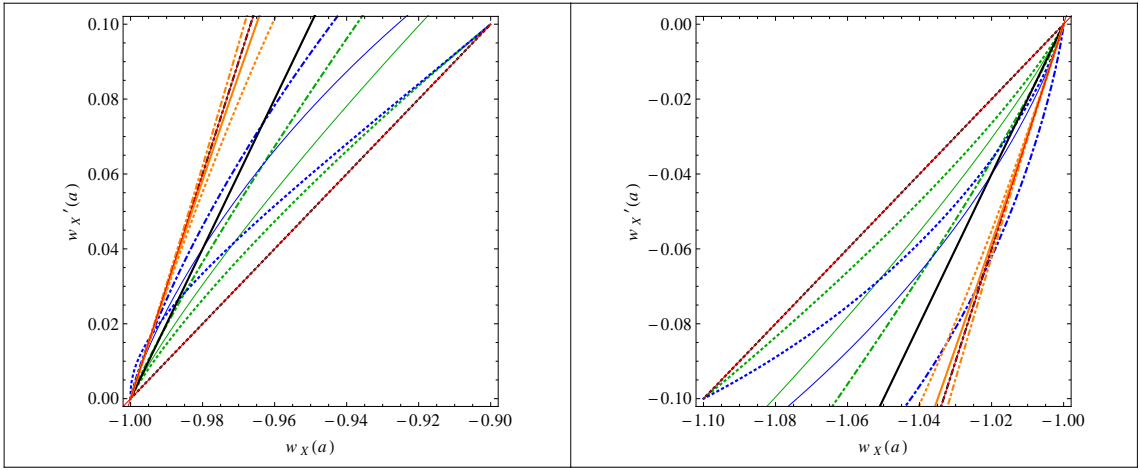}
\caption{Plots depicting generalized
thawing EOS in terms of $w_X-w_X'$ plane for different parameter
values as obtained from theoretical predictions.
The left figure is for $w_X^0=-0.9$ and the right figure corresponds
to $w_X^0=-1.1$.
Black (orange) plots are for CPL, PNGB (Algebraic) thawing
models for $w_X^0=-0.9$ and $w_X^0=-1.1$. The dotted, solid and dot-dashed black lines
are for $c=1$ (CPL) and $c=F=2,\,3$ for the PNGB thawing case and for Algebraic thawing
case they are (in orange) for $p=b=1$, $p=b=2$ and $p=b=6$. 
The blue and green curves are for our generalized thawing EOS.
Green (dotted, solid, dot-dashed) lines are for $n=1.2$ 
($c=1,\,1.2,\,1.5$).
Similarly blue (dotted, solid, dot-dashed) lines are for $n=1.5$ 
($c=1,\,1.2,\,1.5$). The area between solid red lines (overlapped with dotted and dot-dashed black lines) is the 
allowed thawing region \cite{cald:8}.}
\label{w-w'example}
\end{figure*}

In order to test the validity of our generalized model we show
in  Fig. \ref{w-w'example}, the theoretically predicted  
$w_X-w_X'$ $\left(w_X'=\frac{dw_X}{d\ln(a)}\right)$ plots for
different thawing models that arise for different values of $n$ and $c$ 
(we will put constrains on this $w_X-w_X'$ plane 
with direct observational data later in this paper).
We find from Fig. \ref{w-w'example}, theoretically obtained $w_X-w_X'$ plane 
for different combinations of $c$ and $n$
in our model satisfy the allowed regions for the same \cite{cald:8}.
In Fig. \ref{w-w'example}, the left plot is for quintessential thawing
with $w_X^0=-0.9$ and the right one is for the case of thawing 
originated in phantom scenarios with $w_X^0=-1.1$. For $n=1$
with $c=1$ (dotted) we get CPL thawing (Eq. (\ref{cpl})) and for $n=1$
with $c=2,\,3$ (solid and dot-dashed respectively) we get PNGB thawing 
(Eq. (\ref{pngb})). The plots in black in Fig. \ref{w-w'example}
indicate these two models in the $w_X-w_X'$ plane. The orange plots 
are for the Algebraic thawing model with $p=b=1$ (dotted lines), $p=b=2$ 
(solid lines) and $p=b=6$ (dot-dashed lines).
The results with higher values of $n$ are shown by the green ($n=1.2$) and
blue plots ($n=1.5$). The dotted, solid and dot-dashed lines in these
cases corresponds to $c=1,\,1.2,\,1.5$ respectively.



\subsection{Theoretical constraints on the models parameters $n$ and $c$}

\begin{figure}
\includegraphics[height=3.2 in, width=3 in, angle=-90]{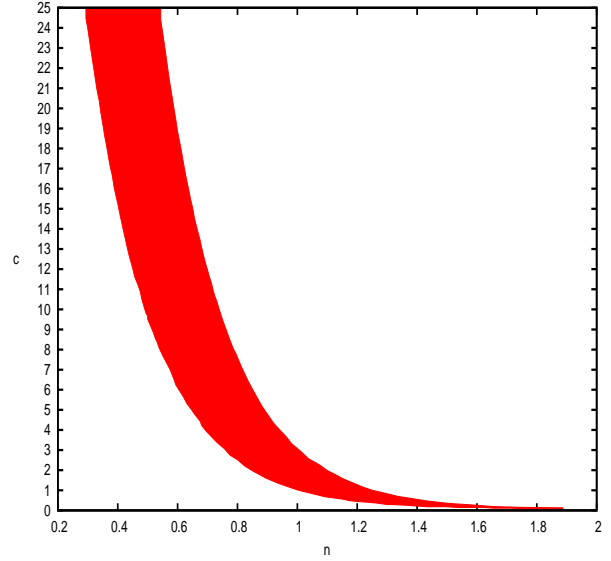}
\caption{Plot showing the theoretical constraints on the models parameters $n$ and $c$ \cite{cald:8}.}
\label{nc}
\end{figure}

In this section we discuss the constraints on the model parameters of our 
genralized thawing dark energy EOS as proposed in the work by Caldwell \etal \cite{cald:8}. 
In Fig. \ref{nc} red region shows the allowed region of the parameter space $(n,~c)$ which is
allowed for thawing dark energy with our generalized EOS. It is also necessary to point out
that our generalized EOS can represent dark energy models other than thawing. The region of 
$(n,~c)$ parameter space except the red zone represents these models. This allowance of $n$ and
$c$ values in Fig. \ref{nc} is also reflected in Fig. \ref{w-w'example}.

%
\subsection{The standard and model independent parameters}
%

As is well-known, any dark energy model must at least probe three parameters directly from observations:

i) the present value of equation of state (EOS) for dark energy  ($w_X^0$)

ii) the present value of matter density ($\Omega_m^0$) 

iii) the Hubble parameter today ($H_0$).

Nevertheless, dark energy model building today is tightly constrained by several 
observations, which, taken together, leave out a very narrow window through
which the model should pass. So, from today's perspectives, apart from the above 
three good old parameters, the supplementary parameters which one needs
to address are the following:


The statefinder pair $\{r,s\}$ \cite{new1} serves as a geometrical diagnostic
to probe the  properties of dark energy in a model independent manner. 
This pair $\{r,s\}$ has been studied extensively in the earlier works 
\cite{new15, new14, new12, new11, new13}. 
For the late universe ($z<10^4$), which is well approximated by the presence of 
matter and dark energy, the statefinder pair $\{r,s\}$ can be expressed
as,
{\small
\begin{eqnarray}
r &=& 1 + \frac{9}{2} \Omega_X w_X (1+w_X) 
- \frac{3}{2} a \Omega_X \frac{dw_X}{da}\label{r}\,\,,\\
s &=& 1 + w_X -\frac{1}{3} \frac{a}{w_X}\frac{dw_X}{da}\label{s}\,\,.  
\end{eqnarray}
}
where $a$ is the scale factor of the universe and $\Omega_X$ is 
the dark energy density parameter. 
In the
late universe we have $\Omega_m+\Omega_X=1$, $\Omega_m$ being the matter
density parameter. 
For $\Lambda CDM$ model, it can be checked that
the statefinder pair $\{r,s\}$ takes the value $r=1$ and $s=0$. Any
deviation in $r$ from $1$ and $s$ from $0$, indicates the existence
of varying dark energy in the universe.

The $Om$ parameter proposed by Sahni \etal \cite{new2}, is another tool to
distinguish the dynamical dark energy from the cosmological constant.
The uncertainty in matter density parameter allows significant  
errors in cosmological reconstructions of dark energy. $Om$
parameter can in practice differentiate between the models,
independent of the matter density parameter. 
The $Om$ diagnostic has been studied well in the earlier works
\cite{new25,new2,new21,new24,new23,new22}.
$Om$ parameter is 
defined in terms of Hubble parameter which can directly be
measured in cosmological observations. The two-point $Om$ \cite{new2} diagnostic is 
given by,
{\small
\begin{eqnarray}
Om(z_2;z_1)&=& \frac{h^2(z_2) - h^2(z_1)}{(1+z_2)^3 - (1+z_1)^3}\,\,,
\end{eqnarray}
}
where $h(z)=H(z)/H_0$. 

It can be easily seen that for cosmological constant $Om(z_1,z_2)=0$ 
and when $z_1<z_2$, $Om(z_1,z_2)>0$ ($Om(z_1,z_2)<0$) represents
the case of quintessence (phantom) \cite{new2}.
This is how $Om$ evaluated at two
different redshifts ($z_1$ and $z_2$) can help in distinguishing the dark energy model. 
Needless to mention
that this procedure is independent of $\Omega_m^0$ and $H_0$.  
The three-point diagnostic $Om3$ \cite{new2} is defined by,
{\small
\begin{eqnarray}
Om3(z_1,z_2,z_3)&=& \frac{Om(z_2;z_1)}{Om(z_3;z_1)}\,\,.
\label{om3}
\end{eqnarray}
}
For $\Lambda CDM$ model $Om3=1$.

Another dimensionless parameter, which is useful for determining
the beginning of cosmic acceleration in dark energy model, is the 
average deceleration parameter $q_{\rm av}$, defined as \cite{new2},
{\small
\begin{eqnarray}
q_{\rm av} &=& \frac{1}{(t_2-t_1)}\int_{t_2}^{t_1} q(t) dt\,\,,
\label{qaveqn}
\end{eqnarray}
}
where $q(t)$ is the deceleration parameter.

We use Eqs. (\ref{r}, \ref{s}, \ref{om3}, \ref{qaveqn})
for evaluating the statefinder pair $\{r,s\}$, $Om3$ and $q_{\rm av}$ 
for the case of our generalization of thawing dark energy model.

Further more, we investigate the growth factor $f$ in the context of
this proposed generalized thawing EOS. For this purpose
we assume the generalized thawing dark energy models proposed here, 
are decoupled from the cold matter sector. This would lead to the effect that the 
galaxy cluster formation is not directly influenced by the 
existence of dark energy. But the presence of dark energy alters the
Hubble expansion rate which affects the growth of inhomogeneities
in the cold matter sector. In the linear regime of matter
perturbations, the evolution of the inhomogeneities are
governed by the relation \cite{Wang:1998gt:12}
{\small
\begin{equation}
\frac{d^2\ln\delta}{d(\ln a)^2} +
\left(\frac{d\ln\delta}{d\ln a}\right)^2 +    
\frac{1}{2}\left(\frac{d\ln\delta}{d\ln a}\right)(1-3w_X(1-\Omega_m))=\frac{3}{2}\Omega_m
\label{Ge}
\end{equation}
}
where $\delta=\delta\rho_m/\rho_m$ is the matter density contrast with 
$\rho_m$ being the matter density. The growth factor $f$ is
defined as \cite{Wang:1998gt:12},
{\small
\begin{equation}
f=\frac{d\ln\delta}{d\ln a}
\label{Gf}
\end{equation}\,\,.   
}

Eq. (\ref{Ge}) can be written in terms of growth factor $f$
(defined in Eq. (\ref{Gf})) as,
{\small
\begin{equation}
\frac{df}{d\ln a} +
f^2 + \frac{1}{2}f(1-3w_X(1-\Omega_m))=\frac{3}{2}\Omega_m
\label{Ge2} \,\,.
\end{equation}
}

The growth equation can be expressed in terms of the redshift $z$
by the relation $\ln a= - \ln(1+z)$. 
The growth factor is well approximated by the ansatz \cite{Wang:1998gt:12}
{\small
\begin{equation}
f=\Omega_m(z)^\gamma
\end{equation}
}
where $\gamma$ is termed as "growth index". 
The growth factor $f$ is affected by dark energy models via $\Omega_m(z)$.



\section{Compilation of combined datasets}

For the purpose of putting constraints on the generalized
thawing dark energy EOS, we use the latest Supernova Type Ia (SNe Ia)
Data from the Union 2.1 compilation \cite{sn:3i},
Observational Hubble Data (OHD) \cite{ohd:4b,Abraham:13,ohd:4c,ohd:4a,stern2010,Moresco2012,ohd:4d}, Cosmic
Microwave Background Data (CMB) from 9 year WMAP results 
\cite{Bennett2012}
and BOSS data from SDSS-III 
\cite{bao:6b}.
There are a total of 607 data points 
(580 data point from SNe Ia, 25 from OHD, and 1 each from CMB and BOSS).
We  make a combined $\chi^2$ analyses of the data sets comprising of 
all 607 data points to constrain our model parameters
$w_X^0$, $\Omega_m^0$ and $H_0$, as well as to confront with the 
model-independent parameters mentioned in Section 2. This makes our 
analysis robust.


\subsection{Union 2.1 compilation of Supernova Type Ia Data}

Luminosity distance ($d_L$) measurement of distant supernovae
with redshifts $z$ is the first observational data to probe the
current acceleration of the universe and the dark energy
properties as well. The most recent compilation of the 
Supernova Type Ia Data is given by Union 2.1 dataset \cite{sn:3i}. The data is
tabulated in terms of distance modulus $\mu(z)$ with
redshift $z$. The distance modulus can be written as
{\small
\begin{equation}
\mu(z)=5 \log_{10}(D_L(z)) + \mu_0\,\,,  
\end{equation}
}
where $D_L(z)= H_0 d_L(z)$ (speed of light in vacuum is normalized to unity)
and $\mu_0=42.38-5 \log_{10} h$ with $h$ given by 
$H_0=100h\, {\rm Km.Sec^{-1}.Mpc^{-1}}$.

$\chi^2$ of SNe Ia data is given by,
{\small
\begin{equation}
\chi^2_{\rm SN}(w_X^0, \Omega_m^0, H_0) =  \sum_{i}\left [
\frac {\mu_{\rm obs}(z_i) - \mu(z_i; w_X^0, \Omega_m^0, H_0)}{\sigma_i} \right ]^2
 \,\,,
\end{equation}
}

Marginalizing over the nuisance parameter $\mu_0$, one gets
the $\chi^2$ as,  

{\small
\begin{equation}
\chi^2_{\rm SN}(w_X^0, \Omega_m^0)= A - B^2/C\,\,,
\end{equation}
}
where $A$, $B$ and $C$ are given by,
{\small
\begin{eqnarray}
A&=& \sum_{i}\left [\frac {\mu_{\rm obs}(z_i) - 
\mu(z_i; w_X^0, \Omega_m^0, \mu_0=0)}{\sigma_i}
 \right ]^2 \nonumber \\
B&=&  \sum_{i}\left [\frac {\mu_{\rm obs}(z_i) - 
\mu(z_i; w_X^0, \Omega_m^0, \mu_0=0)}{\sigma_i} \right ]
\nonumber\\
C&=&  \sum_{i}\frac {1}{\sigma_i^2}
\end{eqnarray}
}



%
%
\subsection{Observational Hubble Data (OHD)}
%
Measurements of Hubble parameters from 
differential ages of galaxies provide another way to
probe the late time acceleration of the expanding universe. 
Jimenez \etal \cite{ohd:4b} first utilized this idea of 
measuring Hubble parameter through the differential age method.
Simon \etal \cite{ohd:4c} and later 
Stern \etal \cite{stern2010} 
provides the values of the Hubble parameter
in the redshift range $0.1\lesssim z \lesssim 1.8$ and $0.35<z<1$ respectively.
A total of 21 OHD data points are recorded at present in the literature
\cite{ohd:4b, Abraham:13, ohd:4c, ohd:4a,stern2010, Moresco2012}.  
With the data release 7 (DR7) 
from Sloan Digital Sky Survey (SDSS)
Zhang \etal \cite{ohd:4d} provides 4 new values of hubble parameters at different redshifts.
All the 25 OHD data points, used in this work to constrain the model parameters,
are listed in Table \ref{tab-hubdata}.

The $\chi^2$ function for the analysis of
this observational Hubble data can be defined as
{\small
\begin{eqnarray}
\chi^2_{\rm OHD}(w_X^0, \Omega_m^0, H_0) &=& \sum_{i=1}^{15} \left[ 
\frac { H_{\rm obs}(z_i) - H(z_i; w_X^0, \Omega_m^0, H_0)} {\sigma_i} \right]^2 \,\,.
\end{eqnarray}
}

\begin{table}[h]
\begin{center}
\begin{tabular}{c  c  c}
\hline\hline
$z$ & $H(z)$ & $\sigma_H$\\
 & (${\rm km~ sec^{-1}~ Mpc^{-1}}$) & (${\rm km~ sec^{-1}~ Mpc^{-1}}$)\\
\hline
\hline
0.090 & 69 & 12\\
0.170 & 83 & 8\\ 
0.270 & 77 & 14\\ 
0.400 & 95 & 17\\ 
0.900 & 117 & 23\\ 
1.300 & 168 & 17\\ 
1.430 & 177 & 18\\ 
1.530 & 140 & 14\\ 
1.750 & 202 & 40\\ 
0.480 & 97 & 62\\ 
0.880 & 90 & 40\\ 
0.179 & 75 & 4\\ 
0.199 & 75 & 5\\ 
0.352 & 83 & 14\\ 
0.593 & 104 & 13\\ 
0.680 & 92 & 8\\ 
0.781 & 105 & 12\\ 
0.875 & 125 & 17\\ 
1.037 & 154 & 20\\ 
0.24 & 79.69 & 3.32\\ 
0.43 & 86.45 & 3.27\\ 
0.07 & 69.0 & 19.6\\ 
0.12 & 68.6 & 26.2\\ 
0.20 & 72.9 & 29.6\\ 
0.28 & 88.8 & 36.6\\ 
\hline\hline
\end{tabular}
\end{center}
\caption{Hubble parameter ($H(z)$) versus redshift ($z$) data from 
\cite{ohd:4b,Abraham:13,ohd:4c,ohd:4a,stern2010,Moresco2012,ohd:4d}.
Here $H(z)$ and $\sigma_{H}$ are in ${\rm km~ sec^{-1}~Mpc^{-1}}$.}\label{tab-hubdata}
\end{table}



%
\subsection{CMB Shift Parameter Data}
%

CMB shift parameter $R$, to a great extent, is a model independent 
quantity extracted from CMB power spectrum. It is given by
{\small
\begin{equation}
R(z_*)=(\Omega_m H_0^2)^{1/2}\int_0^{z_*} dz/H(z)
\end{equation}
}
where $z_*$ is the redshift value at the time when photons decoupled
from matter in the universe.
$z_*$ can be calculated as (with $\Omega_b$ being the baryon density parameter)
{\small
\begin{equation}
z_*=1048[1+0.00124(\Omega_b h^2)^{-0.738}[1+g_1 (\Omega_m h^2)^{g_2}],
\end{equation}
}
where the functions $g_1$ and $g_2$ read as
{\small
\begin{eqnarray}
g_1&=& 0.0783(\Omega_b h^2)^{-0.238}(1+39.5(\Omega_b h^2)^{-0.763})^{-1}\,\,,\\
g_2&=& 0.560(1+21.1(\Omega_b h^2)^{1.81})^{-1}\,\,.
\end{eqnarray}
}
%

$\chi^2_{\rm CMB}$ is defined as,  
{\small
\begin{equation}
\chi^2_{\rm CMB}(w_X^0, \Omega_m^0, H_0) = \left [ 
\frac {R(z_*, w_X^0, \Omega_m^0, H_0) - R}{\sigma_R} \right ]^2
 \,\,.
\end{equation}
}
From WMAP 9 year results \cite{Bennett2012}, 
we use $R=1.728 \pm 0.016$ at the radiation-matter decoupling redshist $z_*=1090.97$ .



%
\subsection{Baryon Oscillation Spectroscopic Survey (BOSS)}
%


CMASS Data Release 9 (DR9) sample of Baryon Oscillation Spectroscopic Survey (BOSS)
(a part of SDSS-III) provides constraint on 
the dimensionless combination $A(z)=D_V(z)\sqrt{\Omega_m^0 H_0^2}/z$ (independent of $H_0$). 
We use the measured value of $A(z)$ at $z=0.57$ 
($A_{\rm obs}(0.57) = 0.444\pm 0.014$ \cite{bao:6b}) 
to constrain our model parameters space. 

The $\chi^2$ for the BOSS data is defined as
{\small
\begin{eqnarray}
\chi^2_{\rm BOSS}(w_X^0, \Omega_m^0)= \frac{[A_{\rm obs}(0.57)
 - A(0.57, w_X^0, \Omega_m^0)]^2}{0.016^2}\,\,.
\end{eqnarray}
}
%
\subsection{Combined $\chi^2$ analyses}
%

Combining all the datasets from Sections (3.1) - (3.4), comprising of 
altogether 607 data points, the combined $\chi^2$ can be evaluated as:
{\small
\begin{eqnarray}
\chi^2(w_X^0, \Omega_m^0, H_0) = 
\chi^2_{\rm SN}(w_X^0, \Omega_m^0) + 
\chi^2_{\rm OHD}(w_X^0, \Omega_m^0, H_0)\nonumber\\
\chi^2_{\rm CMB}(w_X^0, \Omega_m^0, H_0) +
\chi^2_{\rm BOSS}(w_X^0, \Omega_m^0)\,\, .
\end{eqnarray}
}

In what follows, we minimize this combined $\chi^2$ with the 
observational data sets and search for possible consequences 
by confronting our generalized model directly with observations.

In the case we consider all the dark energy models i.e., thawing 
as well as non-thawing that can arise
from our generalized EOS the total $\chi^2$ will be function
of $n,~c,~\Omega_m^0,~w_X^0,~H_0$ when we consider combined data sets
consisting of SNe Ia, BAO, OHD and CMB Shift parameter data.
Marginalized $\chi^2$ in general is defined as \cite{chi-mar1, chi-mar2},

\begin{eqnarray}
\bar\chi^2(p_s)=-2\ln \int_{\theta_1}^ {\theta_2} \exp \left[-\frac{1}{2}\chi^2(p_s,\theta)\right] d\theta~~,
\end{eqnarray}
where the $\chi^2(p_s,\theta)$ is marginalized over the parameter $\theta$ in the range 
$\theta_1<\theta<\theta_2$.

\section{Data analysis and results}

In this section our primary objective is to make a combined $\chi^2$ 
analysis for our generalized model as proposed in Eq (\ref{ourmodel})
with SNe Ia, OHD, CMB and BOSS data for the evaluation of the parameters
space and their $1\sigma$ and $2\sigma$ confidence level (C.L.) limits.
We further study these cases to compare between the results
for $n=1$ and other values of $n$ with the different values of $c$.
Our results are tabulated in Table \ref{tab2}, \ref{tab3} and \ref{tab4}. 
There are five parameters in this generalized thawing model and they are
$n$, $c$, $w_X^0$, $\Omega_m^0$ and $H_0$. We fix the values of $n$ at 
$1,\,1.5,\,2$ with different values of $c$ so that we can compare 
different thawing models and find the best fit values of other three 
parameters by $\chi^2$ analyses. The results of $\chi^2$ analyses for 
PNGB and CPL models are furnished as Case I below and the $\chi^2$ 
analyses results for other thawing models with $n=1.5$ and $n=2$ are 
presented as Case II and Case III respectively.



\subsection{Standard parameters for different values of $n ~\& ~c$}


\noindent \underline{\bf Case I: $n=1$ (CPL \& PNGB)}
\vskip 0.2 cm

In what follows, we describe the results obtained for CPL and PNGB
cases which can be obtained from the proposed generalization of $w_X$ (Eq. (\ref{ourmodel}))
with $n=1$.
The $\chi^2$ analyses results for $n=1$ with different values of 
$c$ are tabulated in the Table \ref{tab2}. These are the cases of
PNGB ($1<c<3$) and CPL ($c=1$) thawing dark energy models. Here we choose 
the values of $c$ to be $1,\,1.5,\,2$. It is seen from Table 
\ref{tab2} that best-fit results ($w_X^0$) point towards the existence of phantom type
thawing dark energy in the universe. As the parameter $c$ goes on taking higher values
the phantom nature gets enriched i.e., the deviation of $w_X^0$ from $-1$
goes on increasing.
During this change of
EOS ($w_X$), the value of matter density parameter at present epoch
and present epoch value of the Hubble parameter
remain unaltered. 
Also needless to mention
here that the values of total $\chi^2$ remain unchanged as is evident
from Table \ref{tab2}.

\begin{table}
\begin{center}
\begin{tabular}{|c|c|c|c|}
\hline
n  &  c  &    best-fit values of &   Minimum \\
   &     &($w_0$, $\Omega_m^0$, $H_0$) & value of $\chi^2$\\ 
\hline
1  &  1  &(-1.009, 0.28, 70.5)   & 575.6\\
 &&&\\
\hline
1   &  1.5  &(-1.011, 0.28, 70.5)   & 575.6\\
      &&&\\
\hline
1   &  2 &(-1.013, 0.28, 70.5)   & 575.6\\
 &&&\\
\hline
\end{tabular}
\caption{\label{tab2} Best-fit values of parameters and minimum values
of $\chi^2$ from combined $\chi^2$ analyses of SNIa, CMB, OHD and BOSS data
for Case I.}
\end{center}
\end{table}

\begin{figure}
\centering
\includegraphics[height=4.6in,width=3.5in]{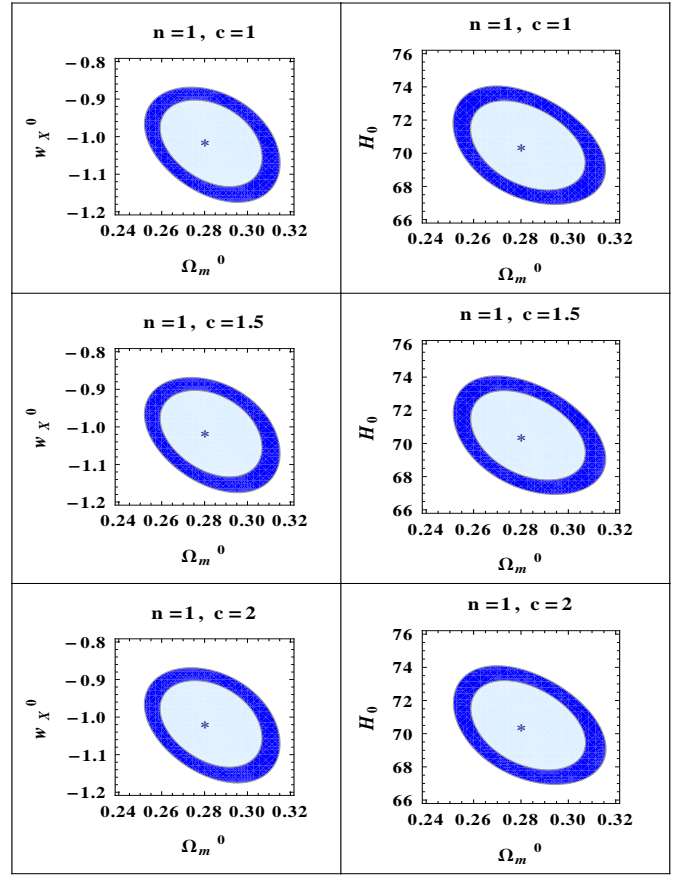}
\caption{This figure shows the contours for the thawing models (CPL \& 
PNGB) arising out of the generalized EOS (Eq. (\ref{ourmodel})) for $n=1$
with different values of $c$ (Case I) as shown in the figure.
$\chi^2$ minimization gives the best fit values which are marked as $*$ in the plot.}
\label{n1}
\end{figure}

In Fig. \ref{n1}, the $1\sigma$ and $2\sigma$ contours of the different
observables e.g., $w_X^0$, $\Omega_m^0$ and $H_0$ for $n=1$ with different
values of $c$ are shown by light blue and dark blue shaded regions 
respectively. The ``$*$'' in the plots represents the best fit values 
obtained by $\chi^2$ minimization (Table \ref{tab2}). Here one can see
that the phantom kind of thawing dark energy is more favoured than the quintessence type
upto $2 \sigma$ C.L. 

\vskip 0.5 cm
\noindent \underline{\bf Case II: $n=1.5$}
\vskip 0.2 cm

Here we investigate the other thawing model that can be originated
for $n=1.5$. The $\chi^2$ minimization results obtained for $n=1.5$ with 
$c= 0.5,\,1,\,1.5$ are tabulated in Table \ref{tab3}. Here also the best-fit
results suggest that the nature of thawing
dark energy is of phantom kind and as $c$ increases the deviation of $w_X^0$ from $-1$ gets increased.
One also sees from Table \ref{tab3} that
the best fit values of present epoch 
matter density parameter $\Omega_m^0$
remain unchanged as the values of $c$ changes. It is also observed that the
best fit values of the Hubble parameters $H_0$ at the present epoch 
also have hardly undergone any changes in these cases. Like the previous
case $\chi^2$ remains unchanged.
 
\begin{table}
\begin{center}
\begin{tabular}{|c|c|c|c|}
\hline
n  &  c  &    best-fit values of &   Minimum \\
   &     &($w_0$, $\Omega_m^0$, $H_0$) & value of $\chi^2$\\ 
\hline
1.5  &  0.5  &(-1.008, 0.28, 70.5)   & 575.6\\
 &&&\\
\hline
1.5   &  1  &(-1.010, 0.28, 70.5)   & 575.6\\
      &&&\\
\hline
1.5   &  1.5 &(-1.012, 0.28, 70.5)   & 575.6\\
 &&&\\
\hline
\end{tabular}
\caption{\label{tab3} Best-fit values of parameters and minimum values
of $\chi^2$ from combined $\chi^2$ analyses of SNIa, CMB, OHD and BOSS data for Case II.}
\end{center}
\end{table}

\begin{figure}
\centering
\includegraphics[height=4.6in,width=3.5in]{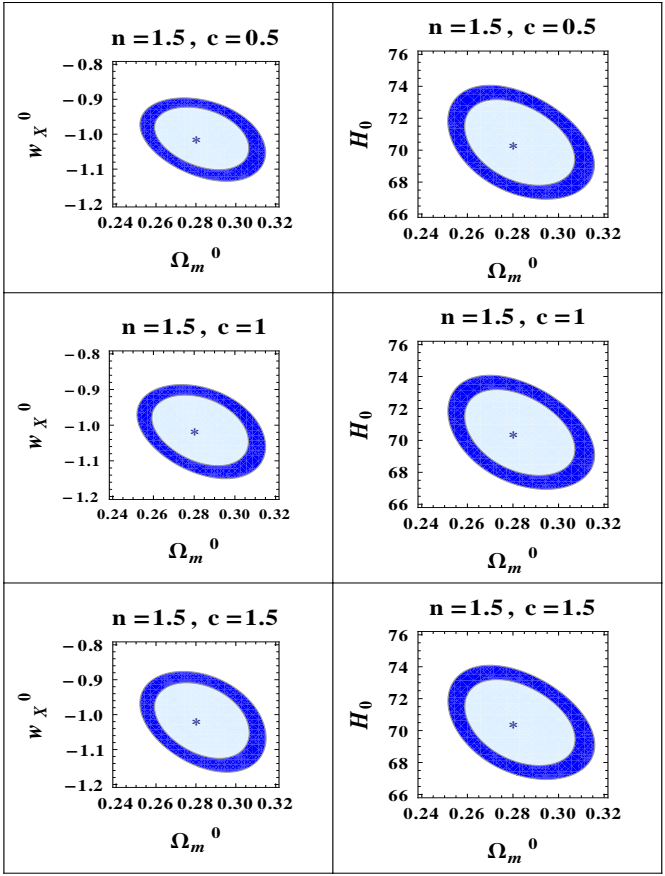}
\caption{This figure shows the contours for the other thawing models
arising out of (Eq. (\ref{ourmodel})) for $n=1.5$ with different 
values of $c$ (Case II) as shown in the figure. $\chi^2$ minimization 
gives the best fit values which are marked as $*$ in the plot.}
\label{n1.5}
\end{figure}
In Fig. \ref{n1.5}, the best fit values (obtained from 
$\chi^2$ minimization) are shown with ``$*$'' symbol and 
the $1\sigma$ and $2\sigma$ contours for different observables
e.g., $w_X^0$, $\Omega_m^0$ and $H_0$ are given by light blue and 
dark blue color shadings respectively. Here one can observe that the phantom type 
of thawing dark energy is more favoured over the
quintessence  upto $2 \sigma$
confidence level.

\vskip 0.5 cm
\noindent \underline{\bf Case III: $n=2$}
\vskip 0.2 cm

Moving onto the $n=2$ thawing scenario, here the results for $n=2$ with
$c=0.5,\,1,\,1.5$ are presented in Table \ref{tab4}.
Like the previous two cases discussed above, it is also evident here 
that the best-fit $w_X^0$ points towards the phantom nature 
of thawing dark energy present in the universe. Also it is seen that $w_X^0$ decreases with the
increasing value of $c$ leaving no significant signatures
in the best-fit values of $\Omega_m^0$ and $H_0$. Also the $\chi^2$ in this case remains unchanged
like the previous two cases.

\begin{table}
\begin{center}
\begin{tabular}{|c|c|c|c|}
\hline
n  &  c  &    best-fit values of &   Minimum \\
   &     &($w_0$, $\Omega_m^0$, $H_0$) & value of $\chi^2$\\ 
\hline
2  &  0.5  &(-1.008, 0.28, 70.5)   & 575.6\\
 &&&\\
\hline
2   &  1  &(-1.011, 0.28, 70.5)   & 575.6\\
      &&&\\
\hline
2   &  1.5 &(-1.013, 0.28, 70.5)   & 575.6\\
 &&&\\
\hline
\end{tabular}
\caption{\label{tab4} Best-fit values of parameters and minimum values
of $\chi^2$ from combined $\chi^2$ analyses of SNIa, CMB and OHD 
and BOSS data for Case III.}
\end{center}
\end{table}

\begin{figure}
\centering
\includegraphics[height=4.6in,width=3.5in]{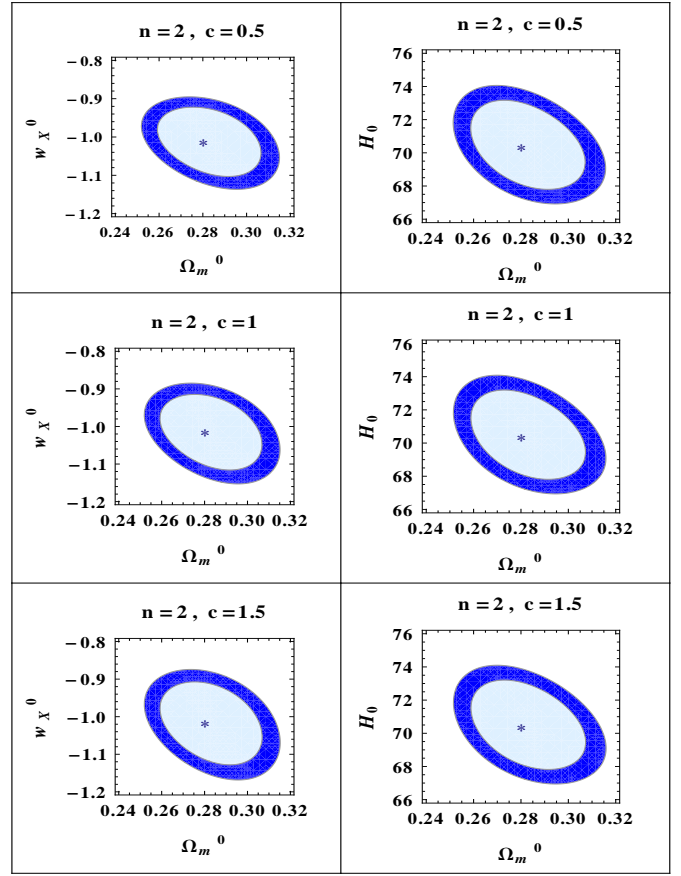}
\caption{This figure shows the contours for thawing 
models (Eq. (\ref{ourmodel})) with $n=2$ for different values of $c$
(Case III) as shown in the figure. $\chi^2$ minimization 
gives the best fit values which are marked as $*$ in the plot.}
\label{n2}
\end{figure}

As in the previous two occasions, best fit values (obtained from $\chi^2$ minimization) 
and  $1\sigma$, $2\sigma$
contours are denoted by ``$*$'' and light blue, 
dark blue color shades respectively in Fig. \ref{n2}. Here also it is easy to figure 
out that the thawing dark energy can be of both quintessence as well as phantom kind (more favoured).
 
Now we compare the results for different values of $n$ with a
particular value of $c$. For $c=1$, one can figure out from the
Tables \ref{tab2}, \ref{tab3}, \ref{tab4} that as $n$ value increases 
from $1$ to $2$, $w_X^0$ shifts from $-1.009$ to $-1.011$ indicating the
enhancement of phantom nature of thawing. 
The present values of matter density parameter $\Omega_m^0$ and Hubble parameter $H_0$ remain
unchanged 
in these cases. The same analogy goes for $c=1.5$ case.
From the above discussions this is apparent that all the three thawing models
(that can be represented by a single form proposed in this work (Eq. (\ref{ourmodel}))) 
produce identical $\Omega_m^0$ and $H_0$ values at least upto 
$2\sigma$ C.L. 

\begin{figure*}
\centering
\includegraphics[height=3in,width=6in]{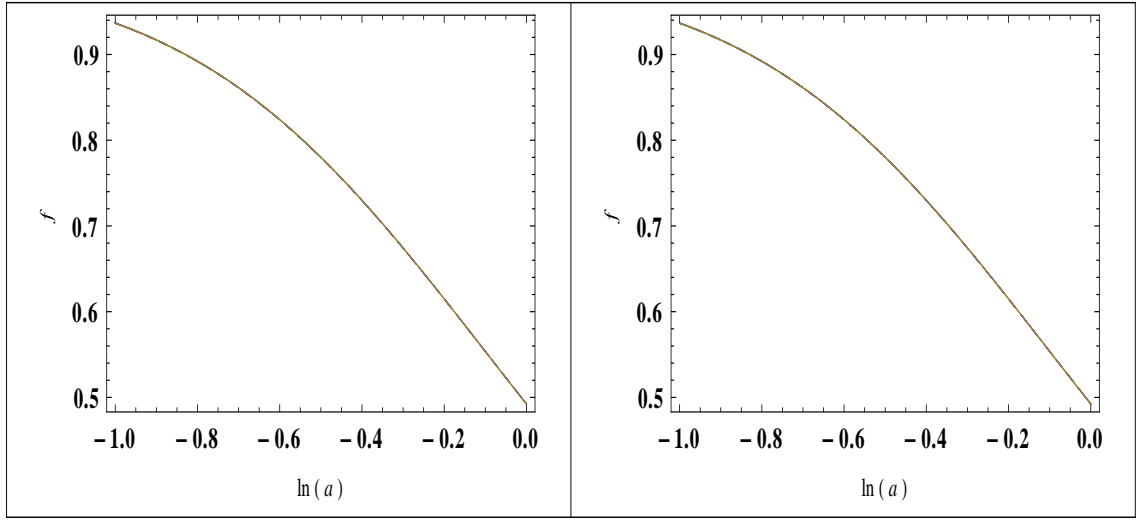}
\caption{Variation of the growth factor with logarithm of scale factor (best-fit plots).
The Left (Right) plot is for the initial condition $f(N=-7)=0.8$ ($f(N=-7)=0.9$) where $N = \ln a$ is the 
number of {\em e-foldings}. Each plot actually depicts 9 different plots overlapping with each other:
plots are for  $n=1$ with $c=1,\,1.5,\,2$ 
and $n=1.5,~2$ with $c=0.5,\,1,\,1.5$.  
}
\label{gp}
\end{figure*}

In Fig. \ref{gp}, the growth factor $f$ is plotted
against the number of e-foldings $N=\log(a)$ for different best fit values of 
$w_X^0$, $\Omega_m^0$ and $H_0$ obtained in the Tables \ref{tab2}, \ref{tab3}, \ref{tab4}.
The left (right) panel is with the initial condition $f(N=-7)=0.8$
($f(N=-7)=0.9)$) for $n=1,~1.5,~2$ with different values of $c$ as described in 
Case I, Case II and Case III in this section. The evolution of the growth factor $f$ is identical in all the
cases suggesting the formation of the same large scale structure
in all cases of thawing considered here (i.e., for CPL and PNGB ($n=1$), Algebraic thawing for $n=1.5$ and $n=2$).
Therefore the growth factor $f$ does not serve as a model discriminator
but acts as a supplementary probe to confirm correct estimation of cosmic
structures formed.

\begin{figure}
\centering
\includegraphics[height=3in,width=3.5in]{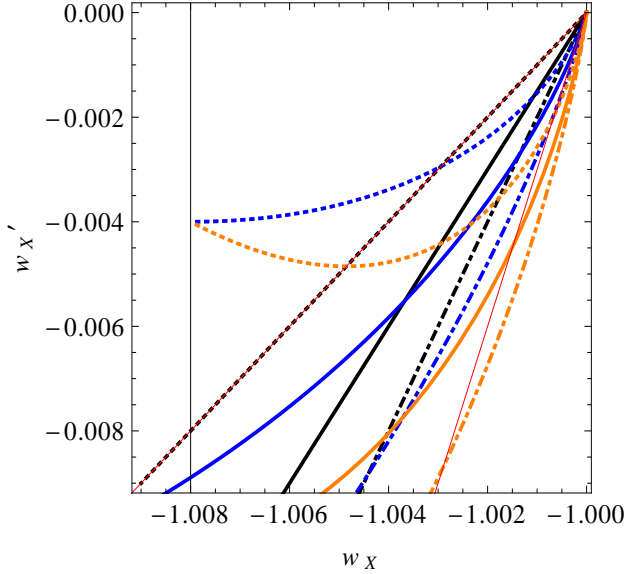}
\caption{
Plot of best-fit $w_X-w_X'$ plane. 
The region between the two red lines is 
allowed $w_X-w_X'$ plane for thawing model \cite{cald:8}.
The black, the blue and the orange lines corresponds to the
thawing models arising out of the generalized EOS (Eq. (\ref{ourmodel})) for
$n=1, ~1.5, ~2$ respectively.
The dotted, solid and dotdashed
lines corresponds to $c=1$, $c=1.5$, $c=2$ for the case of $n=1$ and 
$c=0.5$, $c=1$, $c=1.5$ for the case of $n=1.5$ and $n=2$.}
\label{w-w'1}
\end{figure}

Fig. \ref{w-w'1} depicts the best-fit variation of $w_X'$ with $w_X$ as obtained
using the best fit values of $w_X^0$ from the Tables \ref{tab2}, 
\ref{tab3} and \ref{tab4} for different combinations of $n$ and $c$.
The plots show that we can indeed have non-linear behavior of $w_X$ along with the linear behavior for the 
generalized thawing dark energy model from observations. 
Comparison of these plots with our theoretical predictions, as 
done in Fig. \ref{w-w'example} will be interesting.
Therefore Fig. \ref{w-w'1} goes over Fig. \ref{w-w'example} which was only a 
theoretical prediction. As it turns out from this figure, 
the $w_X-w_X'$ plane indeed serves as a 
model-discriminator for different thawing dark energy models.

\subsection{Model-independent diagnostics}

\begin{figure}
\centering
\includegraphics[height=2in,width=3.5in]{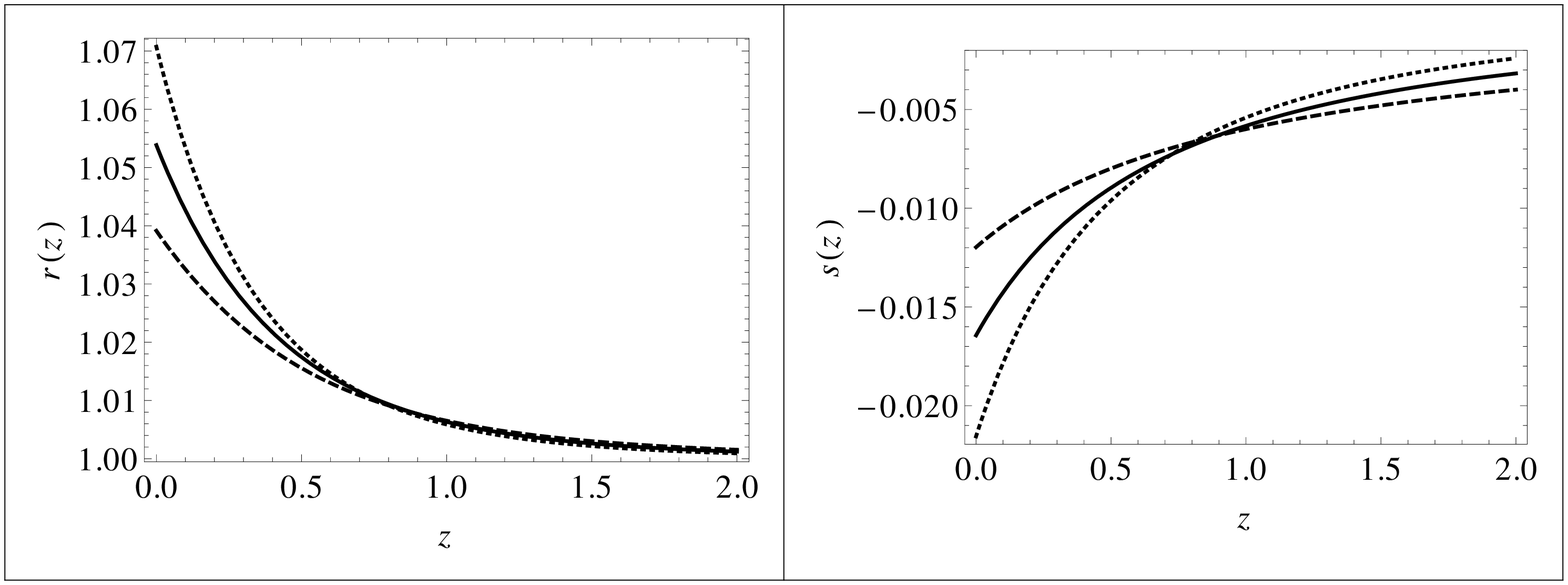}
\caption{Best-fit variations of the statefinders $r$ and $s$ as a function of redshift $z$ (Case I)
for $n=1$ and $c=1$ (dashed), $1.5$ (solid), $2$ (dotted).}
\label{rsn1}
\end{figure}

\begin{figure}
\centering
\includegraphics[height=2in,width=3.5in]{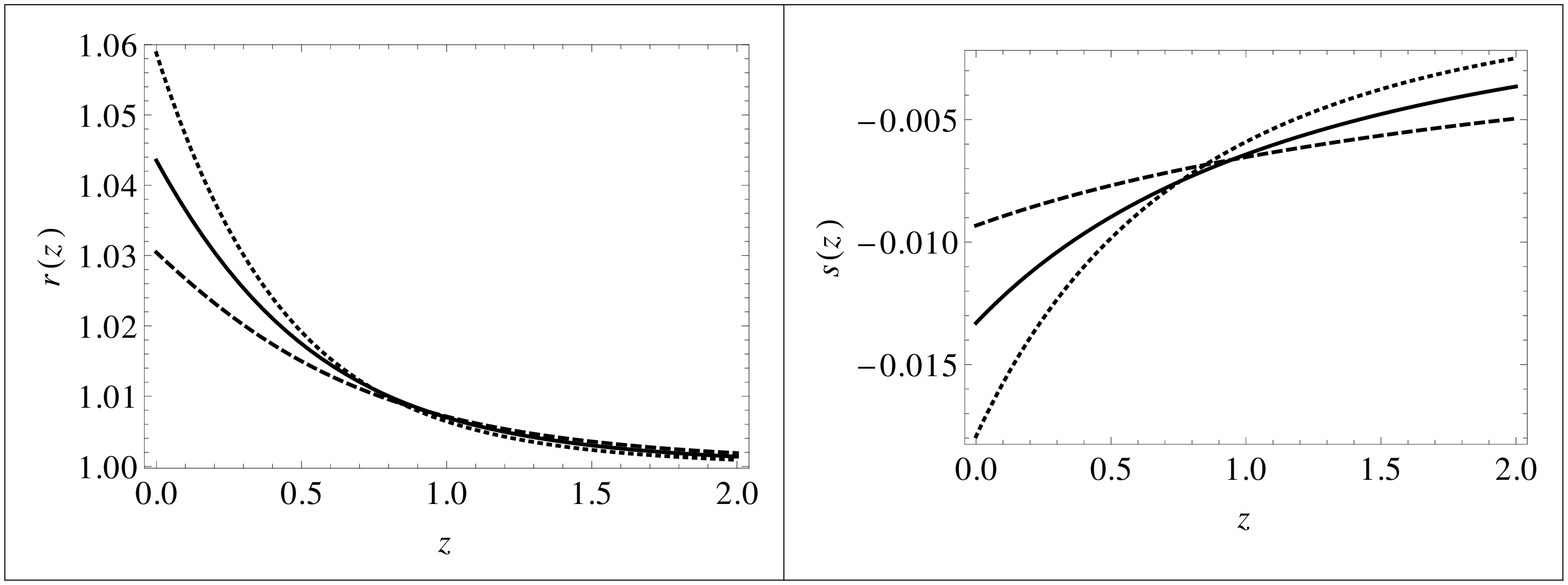}
\caption{Best-fit variations of the statefinders $r$ and $s$ as a function of redshift $z$ (Case II)
for $n=1.5$ and $c=0.5$ (dashed), $1$ (solid), $1.5$ (dotted).}
\label{rsn15}
\end{figure}

\begin{figure}
\centering
\includegraphics[height=2in,width=3.5in]{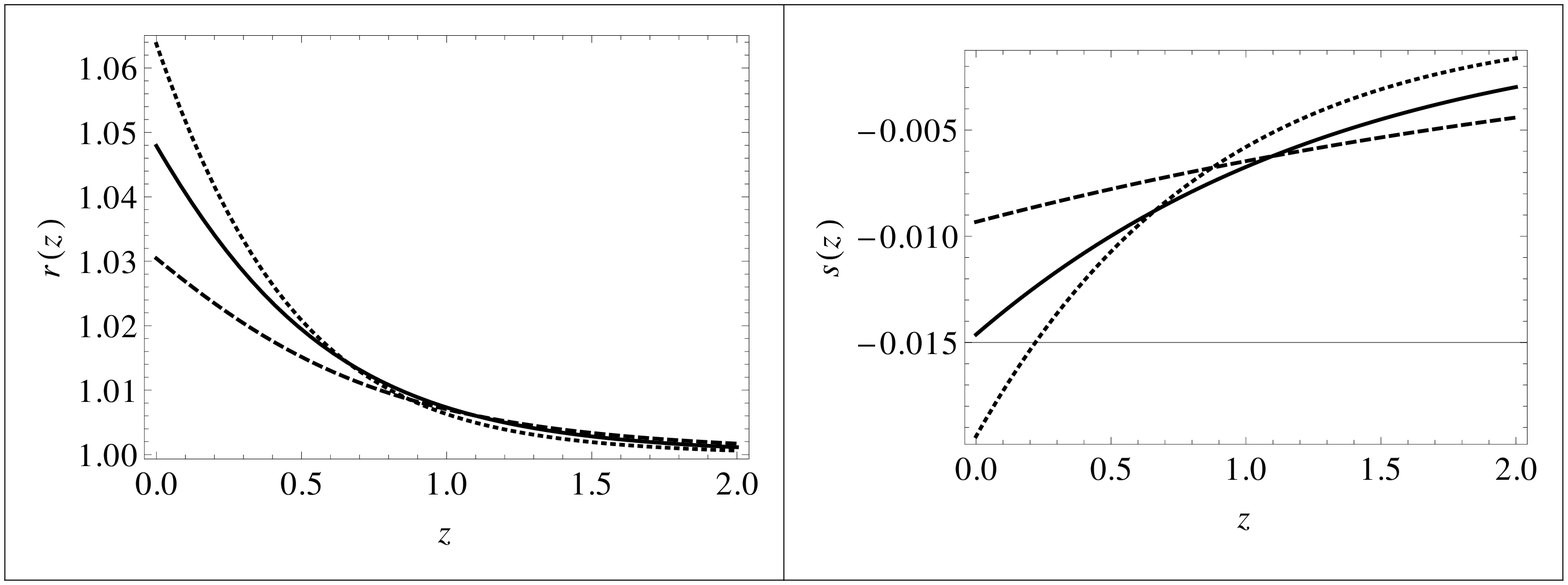}
\caption{Best-fit variations of the statefinders $r$ and $s$ as a function of redshift $z$ (Case III) 
for $n=2$ and $c=0.5$ (dashed), $1$ (solid), $1.5$ (dotted).}
\label{rsn2}
\end{figure}

In Fig. \ref{rsn1} we show the best-fit
variations of the statefinder parameters $\{r, s\}$ with redshift $z$
for $n=1$ case (with the best-fit values of $w_X^0$, $\Omega_m^0$ and $H_0$ 
presented in the Tables \ref{tab2}) 
which 
is known as CPL for $c=1$ or PNGB for other values of $c$.
The dashed, solid and dotted plots are for $c=1,\,1.5,\,2$ respectively.
These plots
bear the clear signatures of thawing as one can see that for higher
values of $z$, the statefinder $r$ tends to $1$ and the statefinder 
$s$ to $0$. This
is because $w_X=-1$ as $z$ increases and since in present epoch
$w_X$ deviates from $-1$, $r$ and $s$ also deviates from $1$ and $0$
respectively. The same features are also observed in the cases of $n=1.5$
and $n=2$ in Fig. \ref{rsn15} and Fig. \ref{rsn2} respectively. 


\begin{figure*}
\centering
\includegraphics[width=6.5in,height= 2.5in]{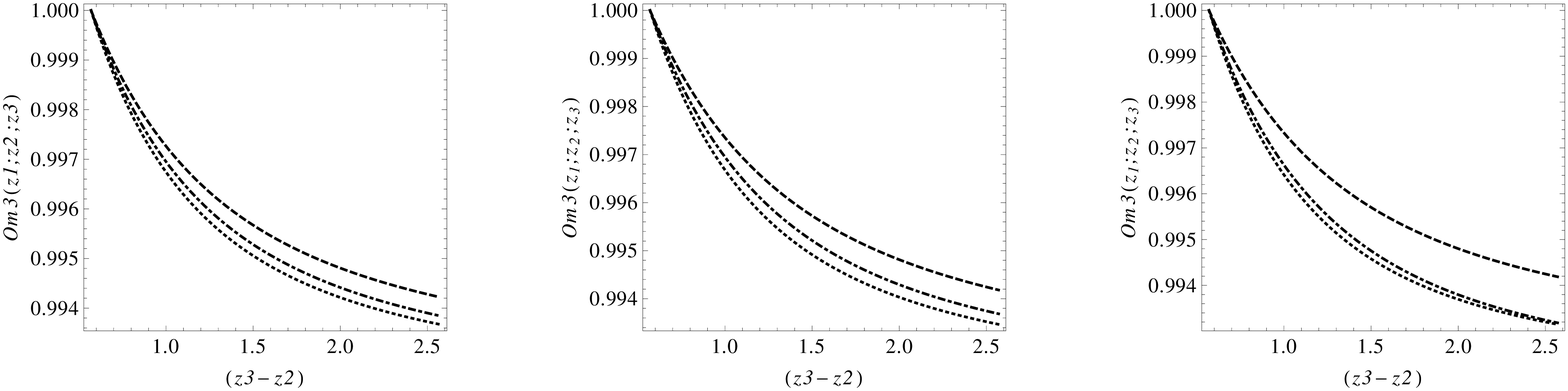}
\caption{The left plot shows the variation of $Om3$ parameter as a function
af redshift $z_3$ with $z_1=0.2$ and $z_2=0.57$ for $n=1$ with
$c=1$ (dashed), $c=1.5$ (dotdashed), $c=2$ (dotted). 
The middle and the right plots shows the same for $n=1.5$ and
$n=2$ respectively with $c=0.5$ (dashed), $c=1$ (dotdashed), $c=1.5$ (dotted).}
\label{Om3}
\end{figure*}

In Fig. \ref{Om3}, we show the best-fit variation of the $Om3$
parameters with the redshift $z_3$ while $z_1$ and $z_2$
are kept at $z_1=0.2$ and $z_2=0.57$ for the best-fit values of 
$w_X^0$, $\Omega_m^0$ and $H_0$ presented in the Tables
\ref{tab2}, \ref{tab3}, \ref{tab4}. 
The plot at the extreme left of Fig. \ref{Om3} shows the variation of $Om3$
parameter for $n=1$ with
$c=1$ (dashed), $c=1.5$ (dotdashed) and $c=2$ (dotted).
The middle and the right plots of Fig. \ref{Om3} show similar variations
for $n=1.5$ and
$n=2$ respectively with $c=0.5$ (dashed), $c=1$ (dotdashed) and $c=1.5$ (dotted).
As $Om3$ is a three point diagnostic, we need three redshift points
to measure its value. We fix two redshift points $z_1$ and  $z_2$
with $z_1=0.2$
\cite{Blake} and $z_2=0.57$ \cite{bao:6b} and allow $z_3$ to be a variable. 
All the variation starts from a point where $z_3=z_2$ that leaves
$Om3=1$ and
the immediate deviation of $Om3$ from $1$ to less than 1 suggests the phantom 
nature of dark energy which is here the 
varying phantom thawing
dark energy.

\begin{figure}
\centering
\includegraphics[height=2.5in,width=3.5in]{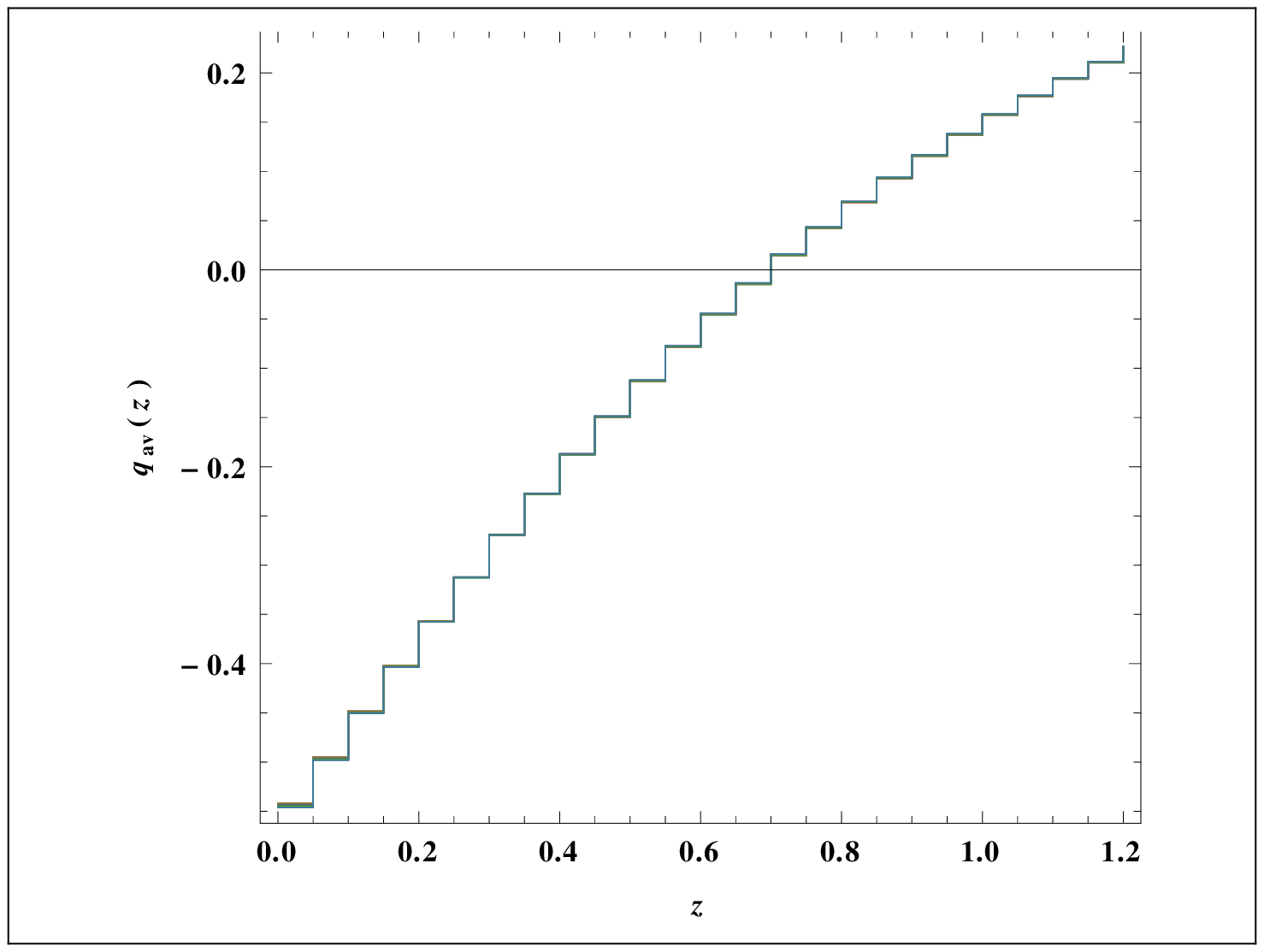}
\caption{Average deceleration parameter $q_{\rm av}$ vs redshift $z$
(best-fit) plot for $n=1, ~1.5, ~2$ with corresponding different values of $c$ (presented in the cases I, II and III).
The plot actually shows all the plots
overlapped on each other.}
\label{qav}
\end{figure}

In Fig. \ref{qav} the best-fit variation of average deceleration parameter 
$q_{\rm av}$ has been plotted with the best fit values of the parameters 
$w_X^0$, $\Omega_m^0$ and $H_0$ obtained in the Tables
\ref{tab2}, \ref{tab3}, \ref{tab4}. It is seen from the plots that all of them overlap
with each other. It is thus evident that average deceleration parameter is not
capable of being a model discriminator, but it does indicate the 
transition period from the deceleration to acceleration phase. In this case
this transition occurs nearly at the redshift $z\sim 7$
as is evident from the best fit plots.

\subsection{Observational constraints on the model parameters}

\begin{figure}
\includegraphics[height=2.5 in, width=2.5 in]{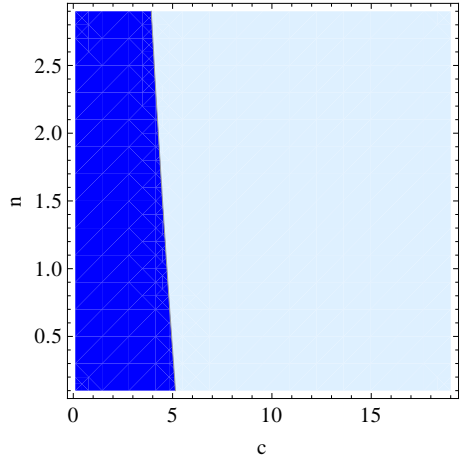} 
\caption{$1\sigma$ and $2\sigma$ contours of $n$ and $c$ marginalized over the 
parameters $\Omega_m^0$ and $w_X^0$.}
\label{ncobs}
\end{figure}

\begin{figure}
\includegraphics[height=2.5 in, width=2.5 in]{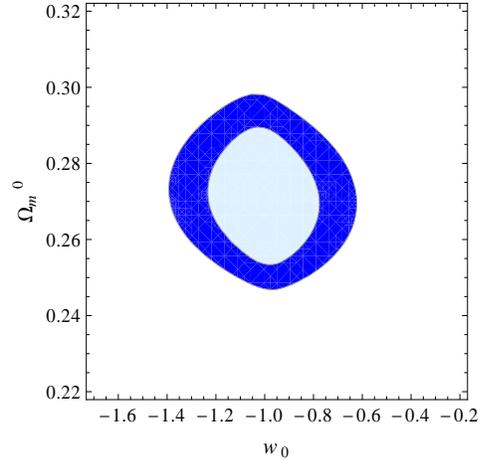} 
\caption{$1\sigma$ and $2\sigma$ contours of $\Omega_m^0$ and $w_X^0$ marginalized over the 
models parameters $n$ and $c$.}
\label{wom}
\end{figure}

In this section we present the result for marginalized contour of $n$ and $c$ (Fig. \ref{ncobs})
with the other parameters $w_X^0$ and $\Omega_m^0$ marginalized over the ranges $-1.7<w_X^0<-0.2$
and $0.1<\Omega_m^0<0.9$.
In Fig. \ref{ncobs} the light blue and the dark blue shades represent
the $1\sigma$ and $2\sigma$ contours.
We also present the marginalized contour of $w_X^0$ and $\Omega_m^0$ (Fig. \ref{wom})
with the model parameters $n$ and $c$ marginalized over the ranges $0.1<n<3$ and $0.1<c<20$.
In Fig. \ref{wom}, the areas enclosed by the smaller inner contour and the bigger outer contour 
represents respectively, $1\sigma$ and $2\sigma$ allowed regions.
In performing so 
the fact that
we have included thawing dark energy models as well as dark energy models which are not thawing,
is evident from the Fig. \ref{nc}. We also would like to mention that in this process
we have used the type Ia supernova data, baryon oscillations spectroscopic survey data and
the cosmic microwave radiation shift parameter data.

The values of $n$ and $c$ leading to thawing dark energy models with our generalized dark energy EOS
(Eq. (\ref{ourmodel})) is described in the subsection II A and II B. From Fig. \ref{ncobs}, one can notice that
present day data does not put any strong constraints on the dark energy models, i.e., claiming that dark 
energy is thawing is not parhaps completely justified from the observational point of view. In other words, 
data does not restricts
us to thawing dark energy models only or present day data is insufficient to favour any particular class of dark
energy models at present.

\section{Discussions and Conclusions}
\label{sec:conclusion}
%
In the present work, we proposed a two parameter generalized EOS, $w_X$,
for thawing dark energy models and studied the dynamics of spatially
flat FRW universe containing radiation, matter and dynamical dark 
energy. This proposal of ours is a minimal generalization of thawing
dark energy EOS and is given by,
{\small
\begin{equation}
w'_X(a)=(1+ w_X)\frac{c}{a^n}\,\,.
\end{equation}
}
This leads to
$w_X(a)=-1+(1+w_X^0) a^c$ for $n=1$ and for other values of $n$,
$w_X(a)=-1+(1+w_X^0)\exp[\frac{c}{(n-1)}(1-a^{(1-n)})]$, where the scheme is that each
value of the parameters $n$ and $c$ defines a specific thawing model,
tuning them will lead to a second model, and so on.
We have also demonstrated that this  minimal generalization
scheme is quite apt as it naturally goes
over the well-known thawing dark energy models such as 
CPL, PNGB and Algebraic thawing, for suitable choice of the two
parameters $n$ and $c$.

We have elaborately discussed the cases with n=1, n=1.5 and n=2
for different values of $c$ ($c=1, 1.5, 2$ for $n=1$ and
$c=0.5, 1, 1.5$ for both the cases of $n=1.5,\,2$). We have shown that though the
parameter $c$ is very important for the slope of the 
$w_X(a)\, vs.\,a$ plot, it barely changes the dynamics of the 
universe. This is quite evident from the average deceleration 
parameter ($q_{\rm av}(z)$) vs redshift $z$ plot (Fig. \ref{qav}),
growth parameter plots (Fig. \ref{gp}) etc and also from the present
values of matter density parameter $\Omega_m^0$ and the Hubble 
parameter $H_0$ as well. 
In this context it is therefore very important to mention that
fine tuning of $c$ does not, at all, effectively change the observables
like the values of density parameters and the Hubble parameter 
at the present epoch 
(vide Tables 
\ref{tab2}, \ref{tab3} and \ref{tab4} for best-fit values and 
Figs. \ref{n1}, \ref{n1.5}, \ref{n2} for $1\sigma$ and $2\sigma$ C.L.s). 
Here it is necessary to mention that in spite of treating the present
epoch value of radiation density parameter $\Omega_r^0$
as a parameter in the numerical analysis, we have chosen its value to be
$\Omega_r^0=5.05\times 10^{-5}$ \cite{Beringer:pdg:2}. This is because of 
the small value of $\Omega_r^0$ which will not change the total density
parameter upto four decimal places and therefore not considering it as a parameter
will not affect density parameters $\Omega_m^0$ or $\Omega_X^0$ significantly.
%

Also we would like to conclude that different values of $n$ would 
lead to same cosmological dynamics for a particular value of $c$
which is evident from average deceleration parameter plot
(Fig. \ref{qav}) and growth factor plots (Fig. \ref{gp}).
These plots clearly demonstrate that it is hardly possible to distinguish between
the results for different
thawing models (related to different values of $n$ and $c$). 
It is necessary here to mention that in calculating growth factor $f$,
we have considered those thawing models that do not modify the Newton's
constant $G$. There exists a class of non-minimally coupled scalar field
models that give rise to thawing and modify the Newton's constant $G$ as well
(see \cite{Ali2012}, \cite{Hossain2012}).
In those cases, no generic form for effective Newton's constant $G_{\rm eff}$
exists as the modification depends on the nature of non-minimal coupling.
Therefore we exclude those thawing models in our proposal of generalized
thawing EOS (Eq. (\ref{ourmodel})).

The analysis thus  reveals a very crucial information about the general class of
thawing dark energy models, namely, different thawing dark energy models can not be distinguished
with the present-day values of matter density, Hubble parameter, the growth factor plots and the average 
deceleration parameter plots.
The importance of our analysis further lies in the fact that this is a quite generic conclusion, 
 since our proposition does take into account within itself almost all the 
thawing dark energy models. So, we claim that one indeed needs to go beyond these parameters
in order to distinguish among thawing dark energy models. These distinguishers come in the form of 
 geometrical
diagnostics like statefinder pair ${r,s}$ and the $Om3$ parameters.
Even though observational data for these parameters are lacking till today,
the analysis succeeds in giving some important predictions which, we believe,
may show a direction of which way to proceed in near future.
These statefinder pair ${r,s}$ and the $Om3$ parameter are shown in 
Figs. \ref{rsn1}, \ref{rsn15}, \ref{rsn2}, \ref{Om3} respectively.

We have also shown in 
Fig. \ref{w-w'1} that the best-fit $w_X-w_X'$ plots can,
as well, serve as another 
discriminator for these thawing models.
Nevertheless, it is also shown in 
Fig. \ref{w-w'1} that $w_X-w_X'$
plots are non-linear for $n=1.5$ and $n=2$. For the
existing thawing models (e.g., PNGB and CPL cases), $w_X-w_X'$
plots are strictly linear. In a recent work, Ali \etal \cite{Ali:16}
have found this kind of nonlinear $w_X-w_X'$ plots arising
from scalar field models. So our generalization can also
produce them for values of $n$ other than $1$ and they are also favoured
well by the recent cosmological observations. Moreover from 
Fig. \ref{w-w'1}, it can
be noted that as $n$ takes higher values, 
only lower values of $c$ are allowed for thawing 
dark energy models.  

From the Fig. \ref{nc} one can finds the values of $n$ and $c$ that would lead to the
thawing models with our generalized form of dark energy EOS (Eq. (\ref{ourmodel})). Therefore
it easy to note that the values of $n$ studied in Tables \ref{tab2}, \ref{tab3} 
i.e., $n=1$ and some of $n=1.5$ lead to thawing but others (Table \ref{tab4}) are not thawing which
is also reflected in the Fig. \ref{w-w'1}. It leads us to also conclude that the values of 
$\Omega_m^0$ and $H_0$ are same for the thawing as well as the non thawing dark energy models
(Tables \ref{tab2}, \ref{tab3} and \ref{tab4} and the Figs. \ref{n1}, \ref{n1.5}, \ref{n2}).
Therefore the beauty of the parametrization lies in its form which generalizes the thawing models 
as well as includes other dark energy models which gives us the opportunity to study all the models
together in the context of present day observational data.

Also From the Tables \ref{tab2}, \ref{tab3} and \ref{tab4}, one can see that the $\chi^2$ values are
a bit low. This is because the error bars in the data sets namely type Ia supernova data, baryon oscillation
spectroscopy data, hubble parameter data and the cosmic microwave background shift parameter data are 
large with respect to this generalized model and in the definition of $\chi^2$ as the error bars appear
in the denominator, we get the a bit low value of $\chi^2$. If the error bars are reduced in the data sets
better results can be obtained and we hope to have well constraints on the model parameters in 
this generalized model in near future.

As mentioned, this is a minimal generalization with the two parameters
$c$ and $n$ and one boundary condition given by $w_X(z=0)=w_X^0$, 
$z$ being the redshift.
There may exist other generalizations with more than two parameters.
So selection can be made on the basis of Akaike Information
Criterion ($AIC$) \cite{aic} and the Bayesian Information Criterion ($BIC$) \cite{bic} that
are defined as,
\begin{eqnarray}
AIC &=& - 2\ln ({\cal L}) + 2p' \label{aic}\,\,,\\
BIC &=& - 2\ln ({\cal L}) + p' \ln N\label{bic}\,\,,
\end{eqnarray}
where $\cal{L}$ is the maximum likelihood value which is given
by $\exp(-\chi^2_{min}/2)$, $p'$ is the number of model parameters and
$N$ is the number of data points used to find the minimum 
value of the $\chi^2$ denoted by
$\chi^2_{min}$. We show the $\Delta AIC$ and $\Delta BIC$
values in Table \ref{tabAB}.

\begin{table}
\begin{center}
\begin{tabular}{|c|c|c|}
\hline
Model  &  $\Delta AIC$  & $\Delta BIC$\\ 
&& \\          
\hline\hline
CPL  &  0  &0\\
\hline
PNGB   &  2  & 6.39\\
\hline
Our Model& 4 & 12.78\\
\hline
\end{tabular}
\caption{\label{tabAB} A comparative analysis of the values of Information criteria
using combined $\chi^2$ analyses of SNIa, CMB, OHD and BAO data.}
\end{center}
\end{table}
Usually from statistical analysis, it is inferred that the models 
having $\Delta BIC$ in the range $0 - 2$ are 
strongly supported, models with $\Delta BIC > 2$ are moderately supported,
and those with 
$\Delta BIC > 6$ are unsupported from perspective of a given data.
However, in cosmology, with the rapid increase of the number of data points
$N$ (we remind the reader that we have used combined dataset),
Eq (\ref{bic}) shows that the $\Delta BIC$ value is always going to increase
with 
introduction of new model parameter(s) $p'$.
This does not essentially mean that the models with least number of parameters
are always favored by observations, though it may appear to be so.
For example, we know
$\Lambda CDM$ model (with the least number of parameters) \cite{Li:15} 
fits the SNe Ia 
data only in the low redshift region i.e, for $z<<1$, and in this vein, most
of the models 
pay the price just because they have additional parameters, though they, in 
fact fair well with observations. 
On this note it should be mentioned that, as demonstrated in \cite{evidence} 
the above information criteria should better be replaced by Bayesian Evidence
calculation, which
gives a value after integrating over all probable states, and hence, does not
suffer from any such limitations of AIC or BIC.
Hence, nowadays, most of the cosmological models are relying more on Bayesian
Evidence calculation, rather than $\Delta AIC$ or $\Delta BIC$ calculation. 
We hope to address this issue in near future.

We are in the era of precision cosmology. Observational data are improving day by day. But these are the error bars
that the data come with makes the constraints on the models poor. Therefore it is very necessary to reduce the error
bars which can improve the constraints on the model parameters further. We used Type Ia supernova data, Baryon Oscillation
spectroscopic survey data, observational hubble data and the cosmic microwave shift parameter data to constrain the
models parameters. Among all these data supernova data influences the analysis the most i.e., the
constrained parameters space depend on supernova data to a great extent. Supernovae data are not that precise at 
the present moment as it comes with large error bars. 
Improving supernova data can probably give us the improved and satisfactory results in future.

\section*{Acknowledgements} 

We are thankful to Supernova Cosmology Project (Union 2.1) http://supernova.lbl.gov/Union/;
Wilkinson Microwave Anisotropy Probe http://lambda.gsfc.nasa.gov/product/map/current/
and Sloan Digital Sky Survey http://www.sdss.org/
for providing us with the online data.
We also thank S. Das and B. K. Pal for some useful comments.


\end{document}